# On the failure of neutron yield scaling in the Dense Plasma Focus

S K H Auluck
International Scientific Committee on Dense Magnetized Plasmas,
Hery 23, P.O. Box 49, 00-908 Warsaw, Poland, skhauluck@gmail.com

Abstract:

The observed scaling of neutron yield in the Dense Plasma Focus (DPF) as the fourth power of the current in the plasma was the principal driver of the growth of DPF research in its early days. Subsequent discovery of failure of this scaling law was also the principal reason for its abandonment by major laboratories. Attempts to understand this failure of scaling have so far been inconclusive. This letter looks at this failure in the context of the recently introduced Generalized Plasma Focus (GPF) problem and suggests a possible reason that can be experimentally examined using small plasma focus devices. This involves restrictions placed on the drive parameter by conservation laws for mass, momentum and energy. A suggested empirical workaround to the problem of neutron yield scaling failure could also be configured as a method for increasing the pressure range for neutron emission in small DPF devices.

The Dense Plasma Focus (DPF) [1,2,3] attracted interest from large national laboratories in the early days of Controlled Fusion Research mainly because of the observed scaling of its neutron yield as the fourth power of current, even though this neutron yield was known to be from a non-thermal reaction mechanism. There was hope that even this non-thermal fusion mechanism could be scaled up to a very intense neutron source producing neutron fluxes similar to a fusion reactor [4] and even up to a breakeven reactor size [5]. Eventually, many large DPF laboratories encountered failure of neutron yield scaling and discontinued their research [3]. The evolution of DPF as a research discipline [3] was profoundly affected by this realization. Current commercial attempts to develop the plasma focus as an energy producing machine [6-8] and optimistic conceptual engineering studies of plasma focus based space propulsion for interplanetary and deep space missions [9,10] are also adversely affected by the failure of neutron yield scaling.

There have been many inconclusive attempts at understanding the cause of this scaling failure. These can be categorised into three cases [3]:

(1) The pinch current limitation effect [11-14]: A larger capacitor bank does not lead to a larger pinch current because of progressively more imperfect matching between the plasma load and its power source.

(2) Failure of neutron scaling: A pinch is formed and is driven at a high current but fails to produce the expected neutron yield [15,16,17].

(3) Failure to form a pinch: A larger capacitor bank or operation at a higher voltage fails to form a pinch - no current derivative singularity or voltage spike results anywhere during the discharge [3,18].

No definitive understanding of the causes for the failure of neutron yield scaling and possible workaround have resulted so far. This is partly due to [3] absence of interpretable data in the corresponding operating regime. The mechanism responsible for this scaling has been mentioned as an open question in a recent comprehensive review [3].

A significant clue to neutron yield scaling issue is provided by the observation [19,20,21] that the "drive parameter" $\mathbb{D} \equiv I_0 / a\sqrt{\rho_0}$ is nearly constant for neutron optimized devices over the energy range of 1 MJ to 0.1 J.

This Letter demonstrates that the constancy of the drive parameter is a direct consequence of conservation laws for mass, momentum and energy. It also provides a reason for the existence of the neutron scaling law and derives its fourth-power-of-current form. For this purpose, it uses a newly proposed [22] scaling theory of the plasma focus "…that is well-grounded in physics, in conformity with existing experimental knowledge and applicable to experimentally untested configurations." This scaling theory is a component of the "Generalized Plasma Focus (GPF) problem" which "…concerns a finite, axisymmetric plasma, driven through a neutral gas at supersonic speed over distances much larger than its typical gradient scale length by its azimuthal magnetic field while remaining connected with its pulse power source through suitable boundaries." This development is a conceptual outgrowth of the Resistive Gratton-Vargas model [23-27], largely retaining its mathematical structure but redefining its basic assumptions.

A generic description is provided in the GPF problem for a class of phenomena that are similar to the plasma focus in having an initial neutral gas ambience, an initial plasma formation phase, a plasma propagation phase and an "approach-to-the-symmetry-axis" phase. The plasma propagation problem for this class of phenomena is decomposed into two weakly interdependent subproblems in the GPF approach [22]. This is achieved by expressing equations governing plasma dynamics – " which could be any variety of single or multi-species MHD fluid model" – into a dimensionless form by writing every descriptive field variable as "a product of a scaling parameter, denoted with a subscript 0 (except the scale length 'a' [of the physical problem]) and a scaled variable, denoted by an overtilde; for example, mass density $\rho \equiv \rho_0 \tilde{\rho}$; radial and axial coordinates $r \equiv a\tilde{r}$; $z \equiv a\tilde{z}$; gradient operator $\vec{\nabla} \equiv a^{-1}\tilde{\nabla}$;

velocity $\vec{v} = v_0 \tilde{v}$; magnetic field $\vec{B} \equiv B_0 \tilde{B}$, electric field $\vec{E} = E_0 \tilde{E}$, pressure $p = p_0 \tilde{p}$, current $I(t) = I_0 \tilde{I}(\tau)$."

The scaling parameter for magnetic field is chosen to be the magnitude of the azimuthal magnetic field produced by a straight current-carrying conductor at a certain radial distance, that for coordinates is chosen to be the physical scale length 'a' of the problem and that for mass density of plasma as the mass density $\rho_0$ of the neutral gas medium. This leads to a unique scaling scheme in which the dimensionless time $\tau$ turns out to be proportional to the charge flow in real time t:

$$\tau = Q_m^{-1} \int I(t) dt \, ; \, Q_m = \pi \mu_0^{-1} a^2 \sqrt{2\mu_0 \rho_0} \tag{1}$$

The scaling parameters for magnetic field, velocity, electric field and pressure are found to be

$$B_0 = \frac{\mu_0 I(t)}{2\pi a \tilde{r}} \, ; \, v_0 \equiv \frac{B_0}{\sqrt{2\mu_0 \rho_0}} \, ; E_0 = v_0 B_0 \, ; p_0 \equiv \frac{B_0^2}{2\mu_0} \tag{2}$$

All the information about the physical device is seen to reside in the scaling parameters. The scaled plasma variables are left without any connection with the physical device.

The first subproblem consists of determining the evolution of an imaginary traveling surface of rotation $\psi(t,r,z) = 0$, called the Gratton-Vargas (GV) surface, whose normal velocity equals $v_0$, the scaling parameter for velocity. This surface belongs to the family of integral surfaces of the Gratton-Vargas (GV) equation [23-27] which obey algebraically defined initial and boundary conditions [23,27]. A general procedure for numerical calculations of this family of travelling integral surfaces is available [23,27]. The resulting data ~~is~~ are used to obtain the inductance associated with a current assumed to be localized at the GV surface. A circuit equation is then used to determine the current as a function of time, determining all the scaling parameters.

This first subproblem uses a result from an earlier Letter [28] regarding a lower bound placed on the scaling velocity by the laws on conservation of mass, momentum and energy. The present Letter revisits this result, which is really a part of the second subproblem described below, and discusses its significance in the context of the failure of neutron yield scaling.

The second subproblem consists of solution of equations governing plasma dynamics in dimensionless form. These equations contain no information about the physical device and their solutions must therefore be universally applicable to the entire class of devices which conform to the generic description and the physics model adopted. In particular, this implies that the physics involved in the existence of the neutron yield scaling must be explainable in terms of the scaling theory and the physical process leading to the failure of the scaling law ought to be observed also at lower scales of energy input.

The interdependence between the two subproblems can be neglected in a first iteration for the fully dissociated and ionized current-carrying zone of the plasma. The magnetic Reynolds number for this case can be high enough to produce a narrow spatial width of the current distribution. Under this condition, the local curvature of the traveling plasma, an aspect related to the first subproblem, can be neglected, assuming the plasma to be locally planar. In a local, orthogonal, curvilinear frame of reference [29] whose axes are aligned with the local tangent to the GV surface, with the azimuthal direction and with the local normal to the GV surface, a traditional planar shock wave solution can therefore be considered as a good first approximation.

Equations for conservation of mass, momentum and energy of the fluid can be written [28] in Hyperbolic Conservation Law form [27] using Maxwell's stress tensor and Poynting's theorem. The equivalent Rankine-Hugoniot jump conditions, with subscripts (0) and (1) denoting quantities on undisturbed and shock-affected sides of the shock discontinuity, can be simplified to obtain [28] the equivalent relations for the Michelson–Rayleigh line

$$\frac{p_{(1)} + B_{(1)}^2/2\mu_0 - p_{(0)}}{V_{(1)} - V_{(0)}} = -\rho_{(0)}^2 s^2 \qquad (3)$$

and the Hugoniot curve

$$\left(\frac{\gamma_{(1)}}{\gamma_{(1)}-1} p_{(1)} V_{(1)} - \frac{\gamma_{(0)}}{\gamma_{(0)}-1} p_{(0)} V_{(0)}\right) - \frac{1}{2}\left(p_{(1)} - p_{(0)}\right)\left(V_{(1)} + V_{(0)}\right) + \frac{B_{(1)}^2}{2\mu_0}\left(\frac{5}{2} V_{(1)} - \frac{3}{2} V_{(0)}\right) + \varepsilon_{d+i} = 0 \qquad (4)$$

where s is the shock speed, $V \equiv \rho^{-1}$ is the specific volume, $\rho_{(0)} = \rho_0$, $B_{(1)} = B_0$ and $\varepsilon_{d+i}$ is the specific energy for complete dissociation and ionization of deuterium (~7.4x10$^8$ J/kg). The polytropic index $\gamma_{(0)} = 7/5$ for the diatomic neutral gas while it is $\gamma_{(1)} = 5/3$ for the fully ionized mono-atomic plasma.

These relations can be reduced to a dimensionless form using the scaling scheme of the GPF with $s \equiv v_0 \tilde{s}$, $\tilde{\varepsilon}_{d+i} = \varepsilon_{d+i}/v_0^2$:

$$\frac{\tilde{p}_{(1)} + 1 - \tilde{p}_{(0)}}{1 - \tilde{V}_{(1)}} = \tilde{s}^2 \qquad (5)$$

$$\left(\frac{\gamma_{(1)}}{\gamma_{(1)}-1} \tilde{p}_{(1)} \tilde{V}_{(1)} - \frac{\gamma_{(0)}}{\gamma_{(0)}-1} \tilde{p}_{(0)}\right) - \frac{1}{2}\left(\tilde{p}_{(1)} - \tilde{p}_{(0)}\right)\left(\tilde{V}_{(1)} + 1\right) + \left(\frac{5}{2} \tilde{V}_{(1)} - \frac{3}{2}\right) + \tilde{\varepsilon}_{d+i} = 0 \qquad (6)$$

Equations (5) and (6) can be reduced to quadratic equations in $\tilde{p}_{(1)}$ and $\tilde{V}_{(1)}$ containing the dimensionless parameters $\tilde{p}_{(0)}$, $\tilde{s}$ and $\tilde{\varepsilon}_{d+i}$. In practice, $\tilde{p}_{(0)} \ll 1$, and can be neglected. Only those values of $\tilde{s}$ and $\tilde{\varepsilon}_{d+i}$ which yield real and positive solutions for $\tilde{p}_{(1)}$ and $\tilde{V}_{(1)}$ can represent a physical plasma. Fig 1 plots the region in the $(\tilde{s}, \tilde{\varepsilon}_{d+i})$ parameter space where both $\tilde{p}_{(1)}$ and $\tilde{V}_{(1)}$ are real and positive for $\tilde{p}_{(0)} \to 0$.

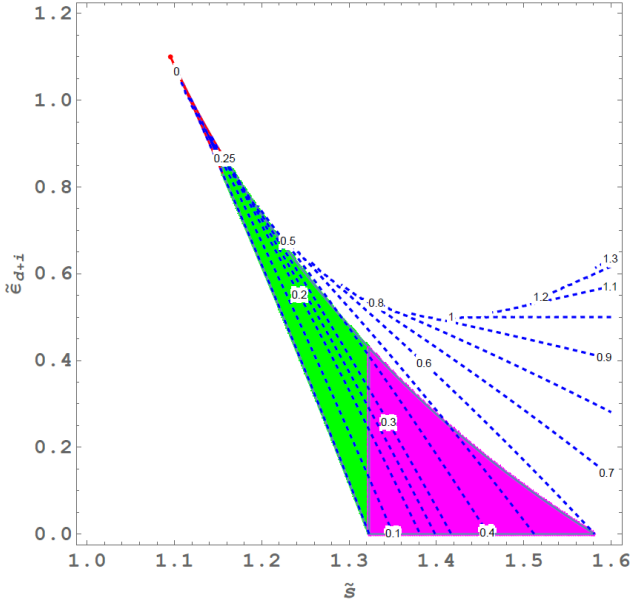

Fig 1: This figure shows the region where both $\tilde{p}_{(1)}$ and $\tilde{V}_{(1)}$ are real and positive for $\tilde{p}_{(0)} \to 0$. The subregion with a red border and blue shading ends in a red dot with coordinates $\left(\sqrt{6/5}, 11/10\right)$. The lower and upper red borders of this subregion are curves described by $\tilde{\varepsilon}_{d+i} = \left(7 - 4\tilde{s}^2\right)/2$ and $\tilde{\varepsilon}_{d+i} = \left(36 - 32\tilde{s}^2 + 9\tilde{s}^4\right)/8\tilde{s}^2$ in the interval $\sqrt{6/5} \leq \tilde{s} \leq 2/\sqrt{3}$. The contiguous subregion shaded in green is bounded between the curves $\tilde{\varepsilon}_{d+i} = \left(7 - 4\tilde{s}^2\right)/2$ and $\tilde{\varepsilon}_{d+i} = \left(5 - 2\tilde{s}^2\right)/2\tilde{s}^2$ in the interval $2/\sqrt{3} < \tilde{s} \leq \sqrt{7}/2$. The next contiguous region shaded magenta is bounded on the lower side by the axis $\tilde{\varepsilon}_{d+i} = 0$ and on the upper side by the curve $\tilde{\varepsilon}_{d+i} = \left(5 - 2\tilde{s}^2\right)/2\tilde{s}^2$ in the interval $\sqrt{7}/2 < s < \sqrt{5/2}$. The blue dashed lines are contours of $\tilde{V}_{(1)}$. The region to the left of the shaded region has $\tilde{V}_{(1)} < 0$ and the one to the right has $\tilde{p}_{(1)} < 0$.

Supersonic propagation is not allowed by the conservation laws until $\tilde{\varepsilon}_{d+i}$, which has a very large value at the beginning of the discharge, decreases below the value represented by the red dot in Fig 1:

$$\tilde{\varepsilon}_{d+i} = \varepsilon_{d+i}/v_0^2 \leq 11/10 \qquad (7)$$

After application of voltage across the plasma focus, a weakly ionized plasma is formed over the insulator [3]. As current builds up, the resulting magnetic pressure detaches the still-developing plasma layer from the surface of the insulator in a subsonic flow [3]. The displacement of the plasma remains negligible compared to its initial radius until the beginning of its supersonic movement away from the insulator. The time interval between the start of current and beginning of the supersonic movement is referred to as the lift-off time $t_L$. During this interval, the circuit behaves essentially as a constant-parameter inductive-capacitive-resistive (LCR) circuit, with inductance $L_0$, capacitance $C_0$, resistance $R_0$ and charging voltage $V_0$, whose current follows the well-known expression

$$I_{LCR}(t) = \frac{I_0}{\sqrt{1-\lambda^2}} \exp(-\gamma_0 t) \sin(\omega_0 t) \equiv I_0 \tilde{I}_{LCR}(t); \quad 0 \leq t \leq t_L \tag{8}$$

$$I_0 = V_0 \sqrt{\frac{C_0}{L_0}}; \quad \gamma_0 = \frac{R_0}{2L_0}; \quad \lambda \equiv \frac{R_0}{2}\sqrt{\frac{C_0}{L_0}}; \quad \omega_0 = \frac{\sqrt{1-\lambda^2}}{\sqrt{L_0 C_0}}; \quad \frac{\gamma_0}{\omega_0} = \bar{\lambda} \equiv \lambda/\sqrt{1-\lambda^2} \tag{9}$$

The scaling velocity given by (2) at the surface of the insulator is

$$v_{0,Ins} = \frac{\mu_0 I(t)}{2\pi a \sqrt{2\mu_0 \rho_0 \tilde{R}_I}} \tag{10}$$

Thus, from (7), the scaling parameter for velocity $v_{0,Ins}$ at the insulator has a lower bound for supersonic propagation of a fully ionized plasma

$$v_{LB} \equiv \sqrt{\varepsilon_{d+i} 10/11} \leq v_{0,Ins} \tag{11}$$

Numerically, $v_{LB} \simeq 2.59 \times 10^4 \, m/s$ for deuterium. At this threshold, the shock velocity has the value $s = \sqrt{6/5} v_{0,Ins} \geq 2.84 \times 10^4 \, m/s$.

These numerical values will be different if the plasma is substantially (say 20-80%), but not fully, ionized and the width of the current carrying zone is smaller but not substantially smaller than the radius of curvature. This can be taken into account by applying a correction factor $f_{LB} \sim 1$ to the right-hand side of (11). Note that the limit represented by the left boundary of the shaded region in Fig. 1 corresponds to $\tilde{V}_{(1)} \to 0$ which implies infinite compression. The right boundary corresponds to $\tilde{p}_{(1)} \to 0$, which implies a very weak shock. Physically realistic values must lie well within these boundaries. These can be obtained only by solving the second subproblem referred above. At present, therefore, the correction factor $f_{LB}$ may be considered an empirical number which may be either slightly less than or slightly more than unity.

Using (11), (10) and (8), the reduced lift-off time $\bar{t}_L \equiv \omega_0 t_L$ is seen to be related to the drive parameter [19] $\mathbb{D} \equiv I_0/a\sqrt{\rho_0}$ via the transcendental equation

$$\tilde{I}_{LCR}(\bar{t}_L) = \frac{\exp(-\bar{\lambda}\bar{t}_L)\sin(\bar{t}_L)}{\sqrt{1-\lambda^2}} = f_{LB}\sqrt{\frac{10\varepsilon_{d+i}}{11}}\frac{2\pi a\sqrt{2\mu_0\rho_0}}{\mu_0 I_0}\tilde{R}_I = f_{LB}\sqrt{\frac{10\varepsilon_{d+i}}{11}}\frac{2\pi\sqrt{2\mu_0}}{\mu_0}\frac{\tilde{R}_I}{\mathbb{D}} \equiv \mathbb{R} \quad (12)$$

A given plasma focus device can be considered optimised with respect to its design parameters if it is ensured that (1) the capacitor bank is fully discharged at the dimensionless time $\tau_p$ when the plasma arrives at the anode centre (2) the current at this moment is the maximum that can be achieved. Referring to the scaling theory of the GPF formalism [19] applied to a classical Mather type plasma focus, this implies the following optimum choice of the operating parameter $\varepsilon \equiv \pi a^2 \sqrt{2\mu_0\rho_0}/\mu_0 C_0 V_0$:

$$\varepsilon_{opt} = \alpha\tau_p^{-1}; \quad \tau_p \equiv 2(\tilde{z}_A - \tilde{z}_I) + 1; \quad \alpha \equiv \exp(-\bar{\lambda}\bar{t}_L)(\cos(\bar{t}_L) + \bar{\lambda}\sin(\bar{t}_L)) \quad (13)$$

where $\tilde{z}_A, \tilde{z}_I$ are the lengths of anode and insulator normalized to anode radius a. It can be demonstrated that scaled current $\tilde{I}(\tau_p)$ has no true maximum with respect to the other dimensionless operating parameter $\kappa \equiv \mu_0 a/2\pi L_0$ for a classical Mather type plasma focus. One can, however, require the fraction of energy converted into magnetic energy at $\tau_p$ to be a certain desired value $\eta_{m,des}$. Ideally, one would like the plasma to have a minimum of kinetic energy when it arrives at the axis in order to have a near-equilibrium pinch, suggesting that all the energy that is not dissipated in the circuit resistance or spent in ionization be converted into the magnetic energy. This criterion, together with (13), translates into an optimum value of $\kappa$

$$\kappa_{opt} = \frac{\exp(-2\bar{\lambda}\bar{t}_L)\sin^2(\bar{t}_L)/(1-\lambda^2) + \alpha^2 - \eta_{m,des}}{\left(\eta_{m,des}\mathfrak{L}(\tau_p) - 2\alpha^2\left(\tau_p^{-1}\mathfrak{M}_0(\tau_p) - \tau_p^{-2}\mathfrak{M}_1(\tau_p)\right)\right)} \quad (14)$$

and an optimum value of the zeroth approximation of the scaled current at $\tau_p$

$$\tilde{I}_{0,opt}(\tau_p) = \sqrt{\eta_{m,des}}\bigg/\sqrt{\left(1 + \frac{\tilde{I}_{LCR}^2(\bar{t}_L) + \alpha^2 - \eta_{m,des}}{(\eta_{m,des} - 2\alpha^2\mathfrak{N})}\right)}; \mathfrak{N} \equiv \left(\tau_p^{-1}\mathfrak{M}_0(\tau_p) - \tau_p^{-2}\mathfrak{M}_1(\tau_p)\right) \quad (15)$$

The dynamic inductance $\mathfrak{L}(\tau)$ and its moments $\mathfrak{M}_0$ and $\mathfrak{M}_1$ in (14) can be expressed [21] as algebraic functions of the scaled geometry of the plasma focus. The number $\mathfrak{N}$ is 0.128 for PF-1000 [3], 0.176 for OneSys [3,17], 0.155 for Gemini [17] and 0.145 for LPP-FF1 [6]. Fig. 2 presents a contour plot of the variation of $\tilde{I}_{0,opt}(\tau_p)$ when the ratio of the circuit resistance to circuit impedance and the drive parameter are varied. For clarity, the data are plotted in terms of the ratio $\mathbb{R}$ on the right-hand side of (12), which is inversely proportional to the drive parameter $\mathbb{D}$ and the ratio $\Lambda \equiv \exp\left(-\pi\lambda/\sqrt{1-\lambda^2}\right)$ of consecutive peaks of the damped sinusoidal short circuit current waveform. The figure is calculated using inductance data $\mathfrak{N} = 0.176$ for the OneSys plasma focus [3,17] and $\eta_{m,des} = 0.7$.

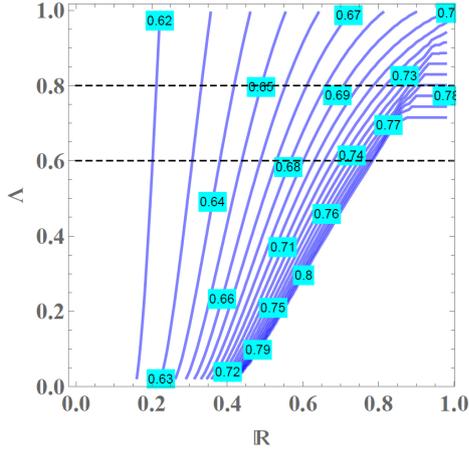

Fig. 2: Contour plot of $\tilde{I}_{0,opt}\left(\tau_p\right)$ in (15). The abscissa is the ratio $\mathbb{R}$ on the right-hand side of (12), which is inversely proportional to the drive parameter $\mathbb{D}$. The ordinate $\Lambda \equiv \exp\left(-\pi\lambda/\sqrt{1-\lambda^2}\right)$ is the ratio of consecutive peaks of the damped sinusoidal short circuit current waveform. The figure is calculated using inductance data $\mathfrak{N} = 0.176$ for the OneSys plasma focus [3,17] and $\eta_{m,des} = 0.7$.

High current capacitor banks are usually designed so that the ratio $\Lambda$ of successive peaks of the damped-sinusoidal short-circuit current falls in the range 0.6 to 0.8, marked by dashed lines in Fig 2. It is seen that the optimum conditions represented by (13) and (14) are satisfied only in a limited range of $\mathbb{R}$. The upper limit on $\mathbb{R}$ (and the lower limit on $\mathbb{D}$) comes from the fact [19] that the function $\exp\left(-\bar{\lambda}\bar{t}_L\right)\sin\left(\bar{t}_L\right)/\sqrt{1-\lambda^2}$ has a maximum value $\exp\left(-\lambda\operatorname{ArcCos}(\lambda)/\sqrt{1-\lambda^2}\right)$. The lower limit on $\mathbb{R}$ (and the upper limit on $\mathbb{D}$) comes from optimization conditions.

This means that the operating value of the drive parameter $\mathbb{D} \equiv I_0/a\sqrt{\rho_0}$ is allowed by the conservation laws to vary only over a limited range of multiples of $\mathbb{D}_0 \equiv f_{LB}\sqrt{10\varepsilon_{d+i}/11}\left(2\pi\sqrt{2\mu_0}/\mu_0\right)\tilde{R}_I$, which equals $2\times10^8 f_{LB}\tilde{R}_I\left(\text{amp}/\text{m}\times\sqrt{\text{kg}/\text{m}^3}\right)$ for D2. This conclusion is verified by plotting these data over the values of $\mathfrak{N}$ in the range 0.1-0.2 and of $\eta_{m,des}$ in the range 0.5-0.8. Outside this limited range, the energy transfer would be sub-optimal. For example, energy may either remain in the capacitor or the plasma may have too much kinetic energy that leads to early disassembly by instabilities.

Condition (13) can be rewritten as

$$\frac{a(\tilde{z}_A - \tilde{z}_I + 1/2)}{\frac{\pi}{2}\sqrt{L_0 C_0}} = \frac{\sqrt{2\mu_0}\alpha \mathbb{D}}{2\pi^2} = \frac{2\alpha}{\pi} f_{LB} \sqrt{\frac{10\varepsilon_{d+i}}{11}} \frac{\tilde{R}_I}{\mathbb{R}} \qquad (16)$$

The numerator on the left-hand side is the effective anode length and the denominator is the ideal quarter cycle time. The right-hand side is proportional to the lower bound $v_{LB}$ prescribed by the conservation laws and given by (11). Equation (16) is a more concrete statement of the well-known practice [21] of ensuring that the discharge reaches the anode centre in coincidence with the ideal quarter cycle time of the capacitor bank. The optimum energy transfer conditions discussed above are equivalent to 'impedance matching' for maximum power transfer in electrical circuits.

The "open question" regarding the origin of the scaling law of fusion yield as the fourth power of current can be answered in terms of the above discussion based on the GPF scaling theory in the following way.

The reaction yield expressed as $\text{Yield} \sim n_{beam} n_{Target} \langle \sigma v_{b-T} \rangle (\text{Vol})(\text{Time})$, with the density of "beam particles" $n_{beam} \sim f_{beam}\rho_0$, of target particles $n_{Target} \sim f_{Target}\rho_0$, velocity-averaged beam-target reaction rate $\langle \sigma v_{b-T} \rangle$, volume of the reaction zone $(\text{Vol}) \sim f_{Vol}a^3$ and reaction time $(\text{time}) \sim f_{time} a/v_{0,pinch}$, can be shown to scale as

$$\text{Yield} \sim f_{beam} f_{Target} \langle \sigma v_{b-T} \rangle f_{Vol} f_{Time} \frac{2\pi\sqrt{2\mu_0}}{\mu_0} \frac{\tilde{r}_{pinch}}{\mathbb{D}_0^{\,5} \tilde{I}_{0,opt}(\tau_p)} I_0^4 \qquad (17)$$

The question of why a universal scaling law for reaction yield must exist is answered by the feasibility of decomposition of the GPF plasma dynamics problem into two weakly interdependent subproblems, one that contains all the device-related information but very little physics and the other that contains all the physics but very little device-specific information, *that must apply to all such devices*.

The fourth power of current scaling law is seen to be a consequence of the above demonstration that devices optimised for energy transfer are forced by conservation laws to operate at a value of the drive parameter confined to a narrow range of multiples of a value $\mathbb{D}_0$ related to material properties of the gas. This scaling of the reaction yield is seen to be applicable to all binary nuclear fusion reactions between gaseous reactants.

The "failure of neutron yield scaling" is an inference drawn from the fact that many large plasma focus installations abruptly stop performing according to expectations above some voltage. This fact can be understood in terms of the above discussion. The device that fails to perform above a certain value of the drive parameter needs to be physically tweaked to comply with the limits on the drive parameter and the generalized optimization criteria discussed above.

The new aspect of this discussion is the proportionality of $\mathbb{D}_0$ with $\tilde{R}_I$, the ratio of the outer radius of the insulator to the anode radius. Relation (17) shows that the reaction yield should vary as the inverse fifth power of the $\tilde{R}_I$. Decreasing $\tilde{R}_I$, which is normally ~1, to a value ~ 0.4 by placing the insulator in the shadow of the anode should increase the yield by *two orders of magnitude* provided all the optimization conditions mentioned above are implemented simultaneously.

These insights from the above discussion can be tested in small plasma focus devices. Measurement of the lift-off time and its correlation with the operating drive parameter and insulator radius according to (12) can be an inexpensive test. The change in the operating pressure range by increasing the insulator radius using an add-on insulator might be another indication to look for. Insulators with outer radius less than the anode radius can be tested for their effect on the operating pressure range. Smaller plasma focus devices can therefore be realistically expected to play a significant role in understanding the failure of neutron emission scaling in large devices, not by making measurements with neutrons but by making measurements on the lift-off time.